\documentclass[a4paper,fleqn,usenatbib]{mnras}
\usepackage{txfonts}
\usepackage[T1]{fontenc}
\usepackage{ae,aecompl}

\usepackage{graphicx, amssymb, amsmath, url}

\title[Performance of $D$-criteria in isolating meteor showers]{Performance of $D$-criteria in isolating meteor showers from the sporadic background in an optical data set}
\author[A. V. Moorhead]{
Althea V. Moorhead$^{1}$\thanks{E-mail: althea.moorhead@nasa.gov}
\\
$^{1}$NASA Meteoroid Environment Office, 
Marshall Space Flight Center, Huntsville, Alabama 35812}

\date{Accepted XXX. Received YYY; in original form ZZZ}

\pubyear{2015}

\begin{document}
\label{firstpage}
\pagerange{\pageref{firstpage}--\pageref{lastpage}}
\maketitle

\begin{abstract}
Separating meteor showers from the sporadic meteor background is critical for the study of both showers and the sporadic complex.  The linkage of meteors to meteor showers, to parent bodies, and to other meteors is done using measures of orbital similarity.  These measures often take the form of so-called $D$-parameters and are generally paired with some cutoff value within which two orbits are considered related.  The appropriate cutoff value can depend on the size of the data-set \citep{1963SCoA....7..261S}, the sporadic contribution within the observed size range \citep{1995EM&P...68..339J}, or the inclination of the shower \citep{2001MNRAS.327..623G}.  If the goal is to minimize sporadic contamination of the extracted shower, the cutoff value should also reflect the strength of the shower compared to the local sporadic background.  In this paper, we present a method for determining, on a per-shower basis, the orbital similarity cutoff value that corresponds to a chosen acceptable false-positive rate.  This method also assists us in distinguishing which showers are significant within a set of data.  We apply these methods to optical meteor observations from the NASA All-Sky and Southern Ontario Meteor Networks.
\end{abstract}

\begin{keywords}
meteors -- meteoroids
\end{keywords}

\section{Introduction}


Quantifying the similarity of meteoroid orbits to each other or to potential parent bodies is a critical first step in many meteoroid dynamics studies.  Establishing a degree of orbital similarity between meteors is necessary in order to identify a new meteor shower, for instance.  Orbital similarity is also presumed to exist between showers and their parent bodies.  Finally, even the sporadic meteor complex largely consists of ``sources" of orbitally similar meteors.  Related bodies may be dissimilar due to meteoroid stream evolution, or they may simply appear dissimilar due to measurement uncertainties.  Conversely, two meteors may resemble each other by chance.  It is therefore important to establish the false-positive rate when using orbital similarity criteria to isolate showers or identify parent body candidates. 


The degree to which two meteoroid orbits are similar is often evaluated using some formulation of a similarity parameter.  \cite{1963SCoA....7..261S} formulated the first $D$-parameter, which is computed from two sets of orbital elements (see Section \ref{sec:dsh}).  $D_{SH}$ and its derivatives are larger for less similar orbits and are therefore sometimes referred to as ``dissimilarity parameters" \citep[e.g.,][]{2001MNRAS.327..623G}.  \cite{1963SCoA....7..261S} designed their initial parameter to be simple to compute and acknowledged that it is possible to formulate valid alternatives.

Alternatives were indeed formulated: one such variation is the Drummond $D$-parameter ($D_D$, Section \ref{sec:dd}).  The Drummond orbital similarity parameter \citep{1981Icar...45..545D} resembles $D_{SH}$ in its general form, but the four terms are scaled to have comparable weight and the last two terms use angular distances rather than chords.  \cite{1993Icar..106..603J} conducted a comparison of these two criteria and concluded that $D_{SH}$ is overly dependent on perihelion distance while $D_D$ is overly dependent on eccentricity.  They proposed the use of a hybrid $D$ parameter that inserts the pericenter term of $D_D$ into $D_{SH}$.  More recently, \cite{1999MNRAS.304..743V} introduced a new $D$-parameter ($D_N$, Section \ref{sec:dn}) that is computed from observed meteor quantities: radiant, velocity, and time.  $D_N$ is constructed such that two of its four components are near-invariant under secular cycling of $\omega$.  


In order to use an orbital similarity parameter, some cutoff value must be adopted within which orbits can be considered related. \cite{1963SCoA....7..261S} modeled the false association rate between meteors by randomizing their nodes; they extracted meteor streams by linking meteors for which $D_{SH} \le 0.2$.  This value was specific to their dataset of 360 meteors; the critical value for single-linkage in larger or smaller data sets could be extrapolated from their value as $0.2 \sqrt[4]{360/N}$. \cite{1971SCoA...12...14L} lowered this slightly to $0.8 / \sqrt[4]{N}$.  \cite{1995EM&P...68..339J} later noted that the appropriate single-linkage cutoff is also a function of the relative percentage of sporadic meteors; \cite{1997A&A...320..631J} computed linkage cutoffs by generating fake meteors using the overall meteor orbital element distributions.  In each case, the single-linkage cutoff is a function of data-set size because it identifies meteors that are more closely related \emph{than average}.

Associating meteors with a known shower or parent body is more straightforward than performing cluster analyses.  Generally, all shower meteors lying within a cutoff value are extracted and smaller datasets simply produce fewer extracted meteors.  \cite{1981Icar...45..545D} compared meteor showers with cometary parent bodies and found that good pairings generally fell within $D_{SH} \le 0.25$ and $D_D \le 0.105$.  Similarly, \cite{1991Icar...89...14D} suggested requiring $D_{SH} \lesssim 0.2$ to 0.25 and $D_D \lesssim 0.1$ to 0.125.  \cite{2001MNRAS.327..623G} tested the performance of $D$-criteria in recovering streams from a large radar data set, quoting cutoff values to recover 70\% and 90\% of a stream as a function of inclination.  No matter how tight a cutoff is applied, however, there is some non-zero chance of including sporadic meteors. 


Galligan modeled the level of sporadic intrusion in meteor stream recovery using two of the above orbital similarity parameters, $D_{SH}$ and $D_N$.  Figures 2 and 3 of \cite{2001MNRAS.327..623G} display CDFs of sporadic AMOR meteors associated with stream searches as a function of $D$-parameter cutoff and inclination.  We follow a similar approach; however, we present the sporadic intrusion for three $D$-parameters and for each individual shower considered rather than for inclination bands.  We find it simpler to model the sporadic intrusion on a per-shower basis and compute the cutoff value for each shower at which the total contamination by sporadic meteors reaches 10\%.


We apply our methods to data from the NASA All Sky Fireball Network \citep{2012pimo.conf....9C} and Southern Ontario Meteor Network \citep{2008EM&P..102..241W}.  Combined, these systems have 25 all-sky meteor video cameras grouped in clusters in the U.S and Canada.  Within a cluster, cameras are separated by 80-150 km; this configuration results in overlapping fields-of-view and permits meteor trajectory triangulation.  Observations are automatically processed by the University of Western Ontario's ASGARD software \citep{2008EM&P..102..241W}, which detects meteors and calculates trajectories.  At the time of writing, these networks have measured trajectories for 27,113 meteors over the past 7 years.

\section{Methods}

We investigate the performance of $D_{SH}$, $D_D$, and $D_N$ by comparing the shower association rate with a modeled false-positive association rate (Section \ref{sec:fpmodel}).  We take the measurement uncertainties into account (Section \ref{sec:errors}), using them to assess the probability that a meteor belongs to a given shower.  We investigate showers in order of decreasing significance in our data set, removing probable shower members with each iteration (Section \ref{sec:extract}).

\subsection{False positive modeling}
\label{sec:fpmodel}

We compute the sporadic intrusion, or false positive rate for shower association, by analyzing the distribution of our orbital similarity parameters relative to shower ``analogs" of our own construction.  These analogs are positioned similarly in Sun-centered ecliptic coordinates, but are offset in time from the shower of interest by $\gtrsim 60^\circ$ in solar longitude.  These analogs are designed such that: [1.] they are offset from the original shower in time, [2.] the region of parameter space encapsulated by an orbital similarity cutoff is similar for these analogs as it is for the original shower orbit, and [3.] both shower and analogs have the same position relative to the sporadic sources in Sun-centered ecliptic coordinates.  The only variations not accounted for by this method are seasonal variations in the sporadic sources and contributions from nearly showers.  We attempt to minimize the influence of nearby showers by working our way from the strongest to the weakest shower, removing each along the way (see Section \ref{sec:extract}).

The exact construction of our shower analogs is as follows.  We compute the shower's radiant in geocentric, Sun-centered ecliptic coordinates: $\lambda_g - \lambda_\odot$ and $\beta_g$.  We construct shower analogs by varying $\lambda_\odot$ from $\lambda_{\odot, shower} + 60^\circ$ to $\lambda_{\odot, shower} + 300^\circ$ in increments of 10$^\circ$.  The value of $\lambda_g$ varies at the same rate as $\lambda_\odot$ so that $\lambda_g - \lambda_\odot$ is kept constant; $\beta_g$ and $v_g$ are also unchanged.  The Sun-centered ecliptic radiant and geocentric velocity define the meteor stream's velocity relative to the Earth at the time of the shower.  We compute the Earth's position and velocity at the given solar longitude in the year 2015 using the DE423 ephemeris.  This allows us to complete the meteoroid stream's state vector at peak activity, which we then convert into orbital elements.  This process yields the full set of shower parameters needed to compute $D_{SH}$, $D_D$, and $D_N$.

For a perfectly circular Earth orbit, the above process would result in shower analogs that have the same orbital elements as the initial shower, aside from $\Omega$, which would vary in the same manner as $\lambda_\odot$.  We therefore model shower association false positives in an approach similar to that of \cite{1963SCoA....7..261S}, who chose to randomize the node for their meteor data.  We note, however, that due to the Earth's orbital eccentricity and the perturbations of the Moon, we produce analogs with values of $q$, $e$, $i$, and $\omega$ that are slightly different from the initial shower orbit.

The $D$-parameters for the shower analogs are combined into individual PDFs as well as a single CDF, the latter of which produces a smoothly increasing function that represents the number of expected false positives within a given cutoff value.  Subtracting this predicted false positive rate from the shower $D$-parameter distribution yields an estimate of the true number of shower members within a given $D$ value.

\subsection{Uncertainty handling}
\label{sec:errors}

\begin{figure}
\includegraphics[width=3.3in]{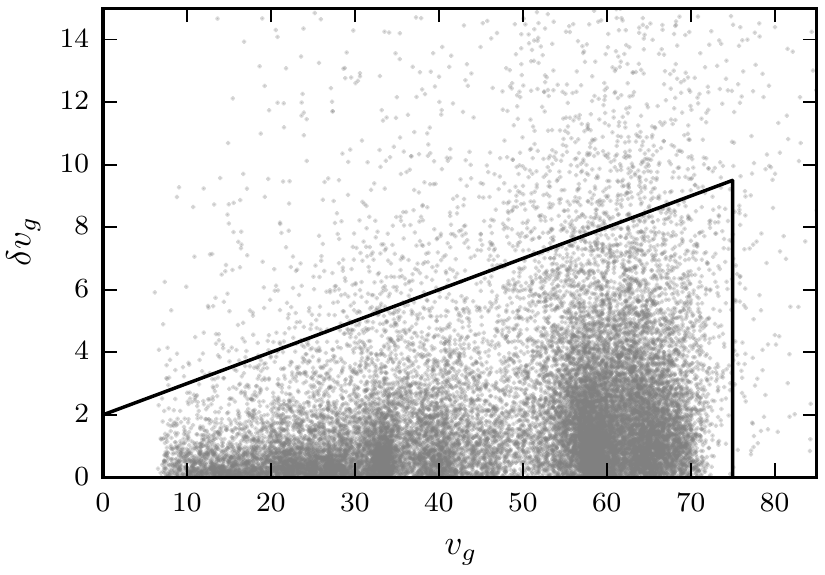}
\caption{Estimated uncertainty in geocentric velocity ($\Delta v_g$) as a function of geocentric velocity for NASA and SOMN meteor trajectories.  Our set of ``quality" meteors lie within the depicted trapezoid.}
\label{fig:vcut}
\end{figure}

\begin{figure*}
\includegraphics[width=\textwidth]{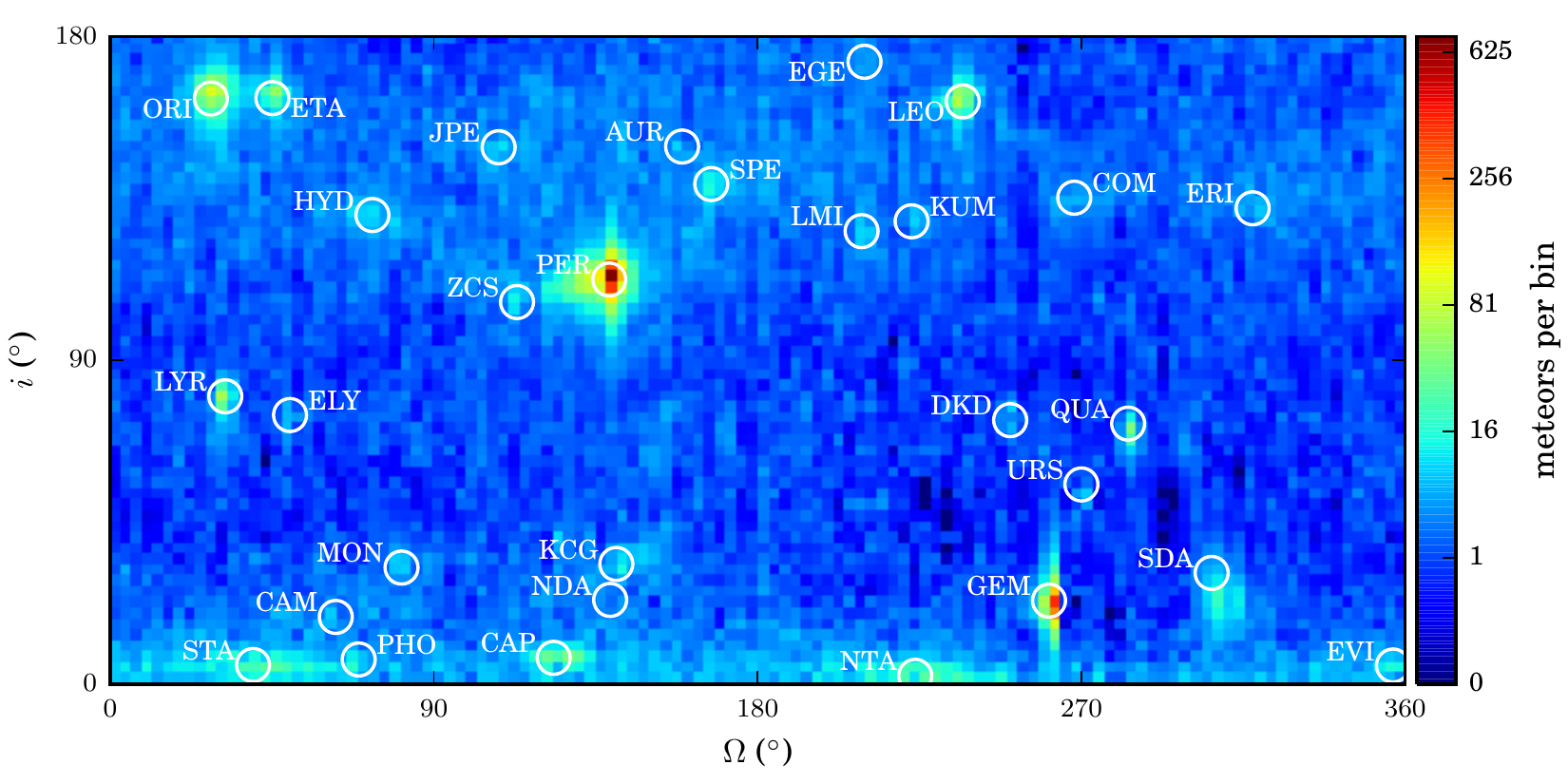}
\caption{Meteor clone ($N=100$) density map in longitude of ascending node ($\Omega$) and inclination ($i$).  Vertical striping occurs because $\Omega$ is generally measured more precisely than $i$.  Blue areas represent low meteor density while red represents high density.  The showers investigated in this paper are circled and labeled with the shower's three-letter code.  Some are clearly visible, such as the Perseids (PER) and Geminids (GEM), while others appear minimal or absent, such as the epsilon Geminids (EGE) and Northern delta Aquariids (NDA).}
\label{fig:heatmap}
\end{figure*}

Each of our meteor trajectory parameters has an estimated associated uncertainty, which can vary substantially between meteors.  To limit the degree to which poorly-constrained orbits affect our results, we take the following measures.  First, we apply some basic quality cuts.  We require that the camera-meteor-camera angle $Q_* \ge 15^\circ$; this geometry requirement favors better-determined trajectories.  We also require that $v_g \le 75$ km/s and that the associated uncertainty $\Delta v_g \le (0.1 \cdot v_g + 1)$ km/s (Figure \ref{fig:vcut}).  This dependence of $\Delta v_g$ on $v_g$ mimics the observed correlation between the two quantities and is intended to avoid biasing the results against faster meteors. 

After making these quality cuts, we create ``clones" of our meteors by assuming Gaussian uncertainties in radiant and velocity (the uncertainty in solar longitude is effectively zero in comparison).  We then compute the orbital elements corresponding to each ecliptic radiant and velocity choice.  These clones are compared to both showers and shower analogs and each clone is individually evaluated for shower membership.  We can then generate a non-binary shower membership probability for each meteor using the sum of its clones.  This Monte-Carlo uncertainty handling produces smoother $D$-parameter distributions and allows us to compute shower membership probabilities that take measurement precision into account.  This step can be omitted in order to analyze the shower association false-positive rate for meteor data that lacks estimated uncertainties.

\subsection{Shower extraction}
\label{sec:extract}

We developed a rough list of showers through the use of an orbital element heat map (Figure \ref{fig:heatmap}).  Showers were selected that correspond to local maxima in this map of ascending node and inclination.  We also analyze several showers that are {\sl not} visible in this heat map for the sake of demonstrating non-detections.  For each shower, we perform our initial $D$ calculations using previously established shower parameters (Table \ref{tab:orbs}).  We compute $D$-parameters for every meteor clone in our data set relative to our nominal shower orbit and radiant.  If the number of low-$D$ meteors exceeds the expected false positive rate, we use the ratio of the false positive $D$ distribution to the shower $D$ distribution to compute the probability that a meteor clone is not a member of the shower as a function of $D$.  We then remove shower members using this distribution, repeating the process for the next strongest shower.

As we extract showers, we re-compute the shower's properties using its members.  In each case, we compute the mean shower orbit as follows:
\begin{align}
\lambda_{\odot,m} &= \arctan{\left({{\langle \sin{\lambda_{\odot}} \rangle}/{\langle \cos{\lambda_{\odot}} \rangle}}\right)} \\
(\lambda - \lambda_{\odot})_m &= \arctan{\left({
\frac{\langle \sin{(\lambda-\lambda_\odot)} \cos \beta \rangle}{\langle \cos{(\lambda-\lambda_\odot)} \cos \beta \rangle}
}\right)} \\
\beta_m &= \arcsin{\left({
{\langle \sin{\beta} \rangle} /
{\| - \vec{u} \|}
}\right)} \\
v_{g,m} &= \langle v_g \rangle
\end{align}
where brackets indicate the arithmetic mean, the appropriate sign is taken for each arctangent computation, and the mean anti-velocity unit vector is $\| -\vec{u} \| = \langle \cos{(\lambda-\lambda_\odot)} \cos{\beta} \rangle \,\hat{x} + \langle \sin{(\lambda-\lambda_\odot)} \cos{\beta} \rangle \, \hat{y} + \langle \sin{\beta} \rangle \, \hat{z}$.  These values are converted to orbital elements, again using the DE423 ephemeris for the Earth to complete the state vector.

\section{Results}

This study accomplishes two tasks.  First, we compare the distribution of $D$-parameters about each shower with that about its analogs to determine whether a shower is detected.  Second, we use the cumulative distribution of $D$-parameters to determine the maximum value of $D$ within which the sporadic intrusion is no more than 10\%.

\subsection{Shower detection}

\begin{figure*}
\includegraphics[width=\textwidth]{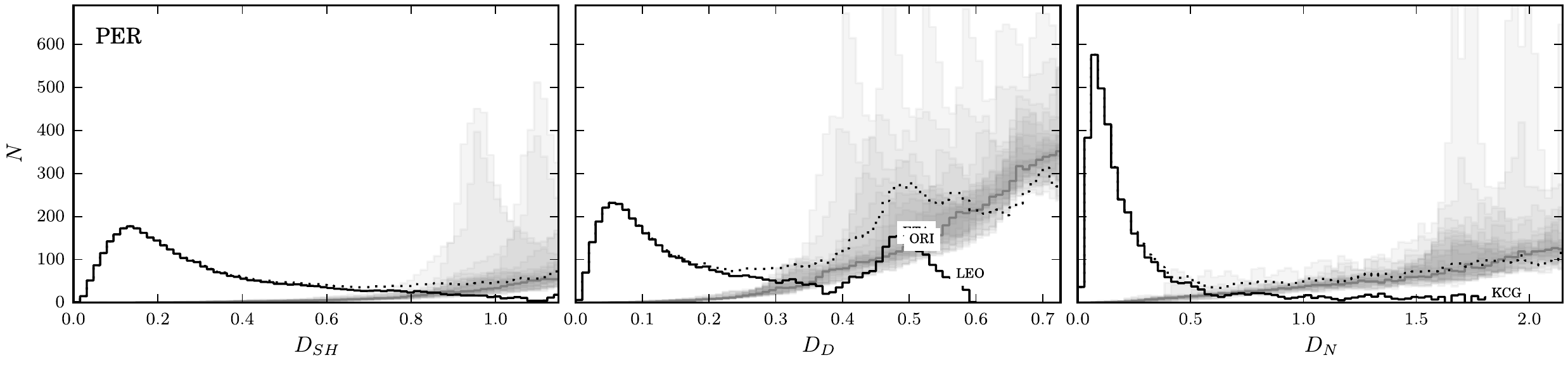} \\
\includegraphics[width=\textwidth]{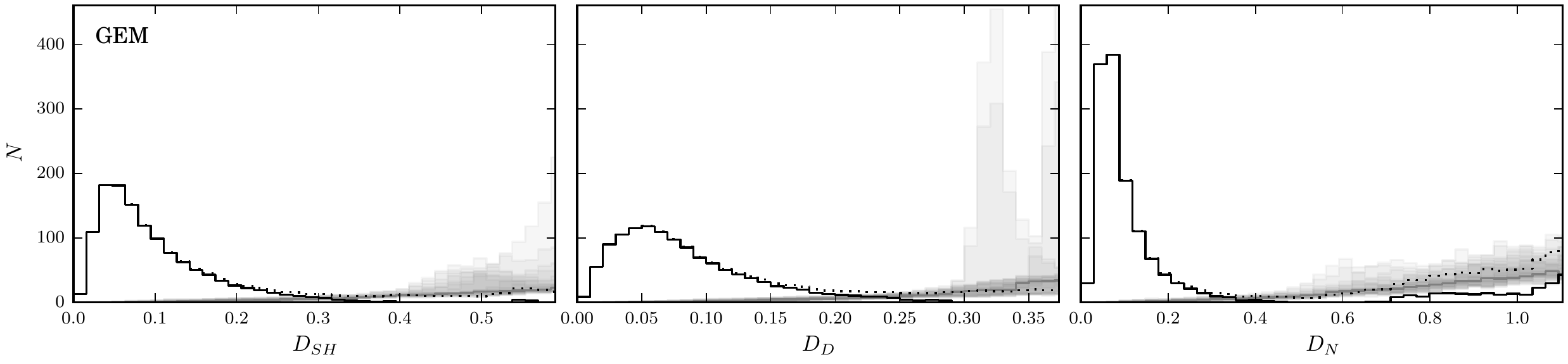} \\
\includegraphics[width=\textwidth]{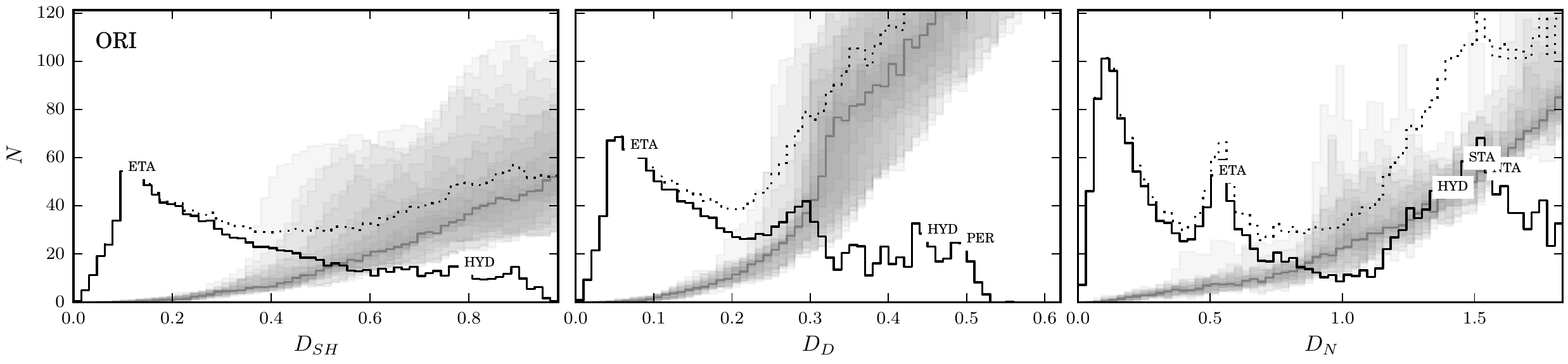} \\
\includegraphics[width=\textwidth]{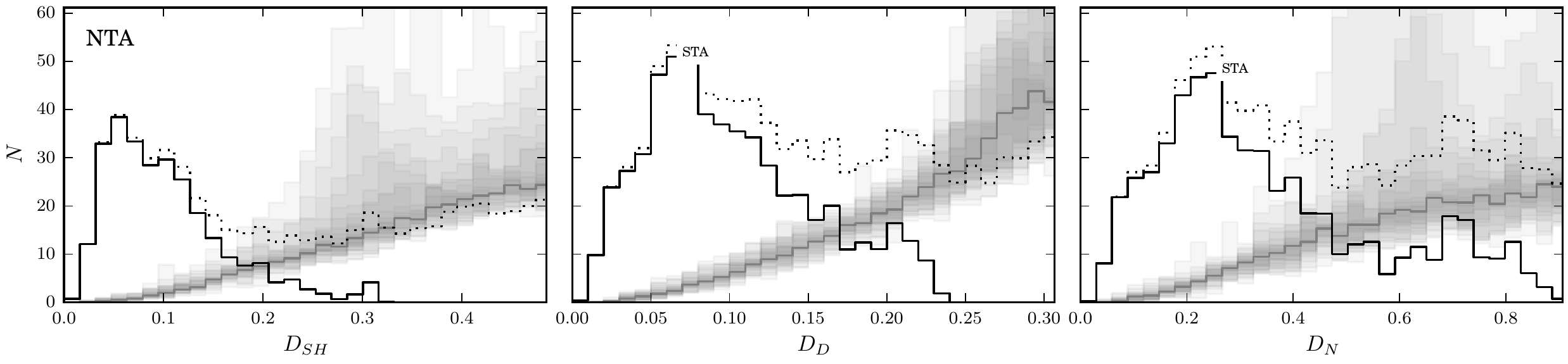} \\
\includegraphics[width=\textwidth]{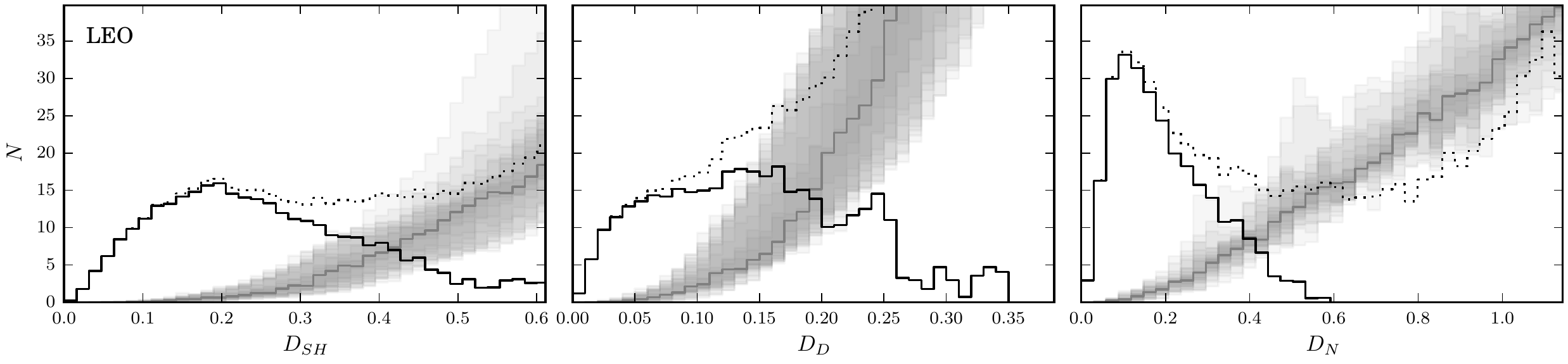}
\caption{The distribution of orbital similarity parameters for five prominent showers in our data set.  Each row corresponds to a single meteor shower, and the columns correspond to the three $D$-parameters investigated: $D_{SH}$ (left), $D_D$ (middle), and $D_N$ (right).  In each plot, the dotted black line depicts the raw distribution of each $D$-parameter computed relative to the shower orbit (Table \ref{tab:orbs}).  The shaded gray area encompasses the distribution of $D$ parameters for each of our 25 ``shower analogs" -- the median appears as a solid gray line.  The solid black line represents the difference between the raw $D$ distribution for the shower and the median $D$ distribution for the shower analogs.  Cases where the $D$-parameter computed between the reference shower and another major shower falls within the depicted range are marked with small shower labels (e.g., ``ETA"). }
\label{fig:big4pdf}
\end{figure*}

Figure \ref{fig:big4pdf} compares $D$-parameter histograms for our data compared to the four most significant showers in our data set (black lines) and the false-positive rate computed from their analogs (gray region).  For each shower, we present histograms for each of our three $D$-parameter choices.  Plot ranges for each shower are synchronized, and each plot domain covers the same fraction of $D_{i, max}$.  

We have also marked the proximity of other major showers in $D$-parameter space.  The September $\epsilon$-Perseids (SPEs), Orionids, $\eta$-Aquariids, and Leonids all lie within $D_D = 0.6$ compared to the Perseids (PER).  The influence of these showers is also reflected in the false positive rate computed using our Perseid analogs.  Note from the middle-top panel of Figure \ref{fig:big4pdf} that the combined contribution of these showers lies within the predicted false positive rate.

Separating the Orionids from the $\eta$-Aquariids is more difficult.  Neither $D_{SH}$ nor $D_D$ are able to distinguish between these two showers; the label marking the $\eta$-Aquariids in Figure \ref{fig:big4pdf} lies on top of the peak of the Orionids.  Two peaks are visible in the $D_N$ distribution, but the $\eta$-Aquariids lie well above the predicted false positive distribution.  This similarity is a result of the two showers being part of the same meteoroid stream -- they should be grouped together by any measure of orbital similarity.  To study these showers separately, it is necessary to slice the stream in two by, for instance, solar longitude.

The Northern and Southern Taurids (NTA and STA) are also challenging to separate using measures of orbital similarity; the only $D$-parameter capable of separating these branches is $D_{SH}$.  Unlike the Orionids and $\eta$-Aquariids, these two showers lie close to one another in both radiant and time of year.  Our solution is to separate them by whether their radiant lies just above or just below the ecliptic.

\begin{figure*}
\includegraphics[width=\textwidth]{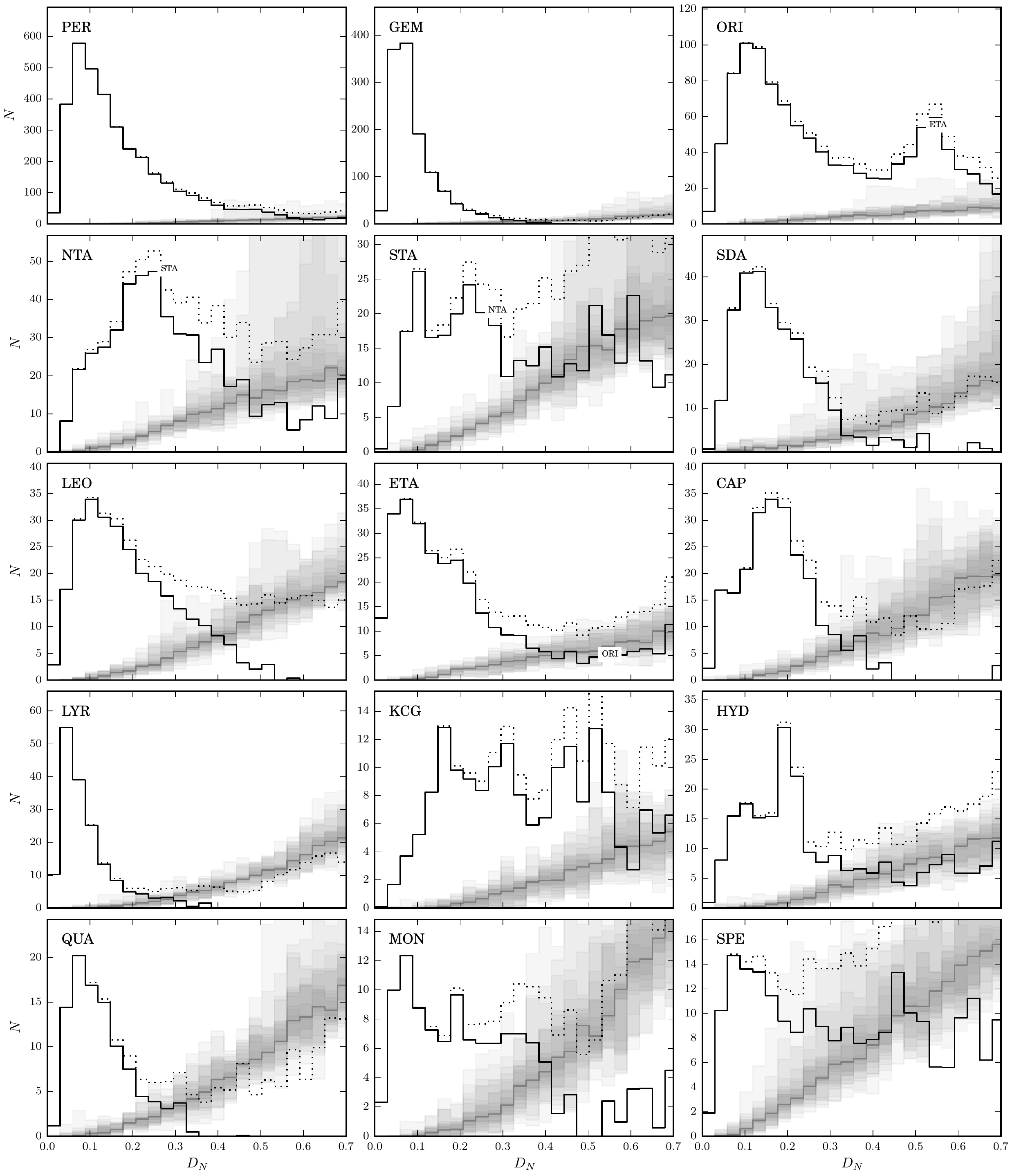}
\caption{The distribution of orbital similarity parameters for the fifteen most prominent showers in our data set. }
\label{fig:pdf15}
\end{figure*}

\begin{figure*}
\includegraphics[width=\textwidth]{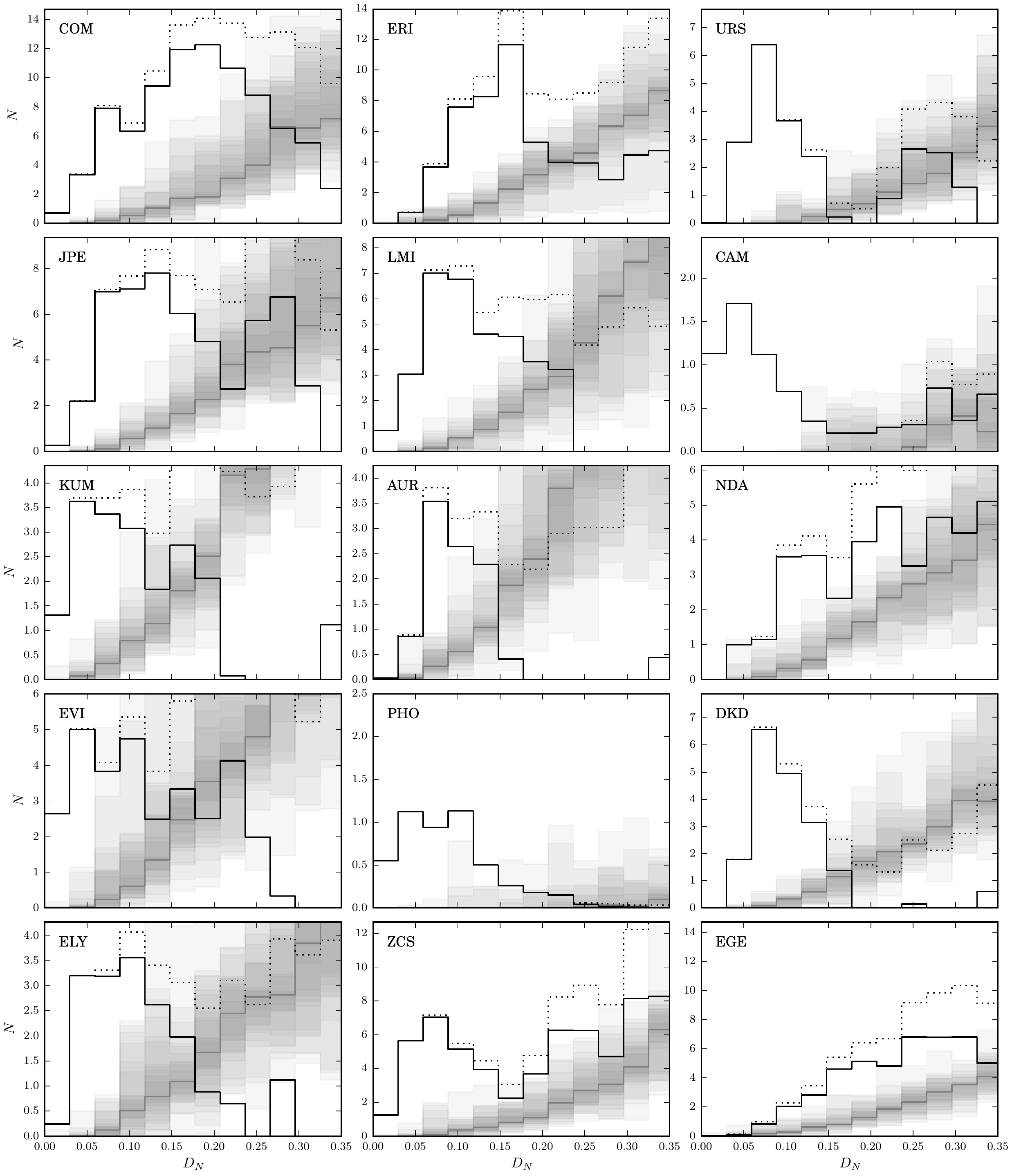}
\caption{The distribution of orbital similarity parameters for fifteen weaker showers in our data set. }
\label{fig:pdf30}
\end{figure*}

The distribution of $D_N$ for each of our 30 showers is depicted in Figures \ref{fig:pdf15} and \ref{fig:pdf30}.  It is obvious from this series of plots that minor showers are overwhelmed by sporadics at lower $D$-parameter thresholds.  Additionally, Figure \ref{fig:pdf30} depicts two showers that we consider non-detections: the Northern $\delta$-Aquariids (NDA) and the $\epsilon$-Geminids.  In both cases, the $D_N$ distribution is hardly distinguishable from our computed false positive rate.

We have also included two minor shower outbursts in this analysis: the May Camelopardalids (CAM) and the December Phoenicids (PHO).  Both of these showers produced a tight cluster of five meteors in our data set that were automatically extracted from our data using the algorithm of \cite{2014JIMO...42...14B} and which corresponded to predicted outbursts.  Figure \ref{fig:pdf30} also indicates that each cluster of 4-5 meteors is anomalously similar compared to the modeled false positive rate.

\subsection{Shower extraction}

\begin{figure*}
\includegraphics[width=\textwidth]{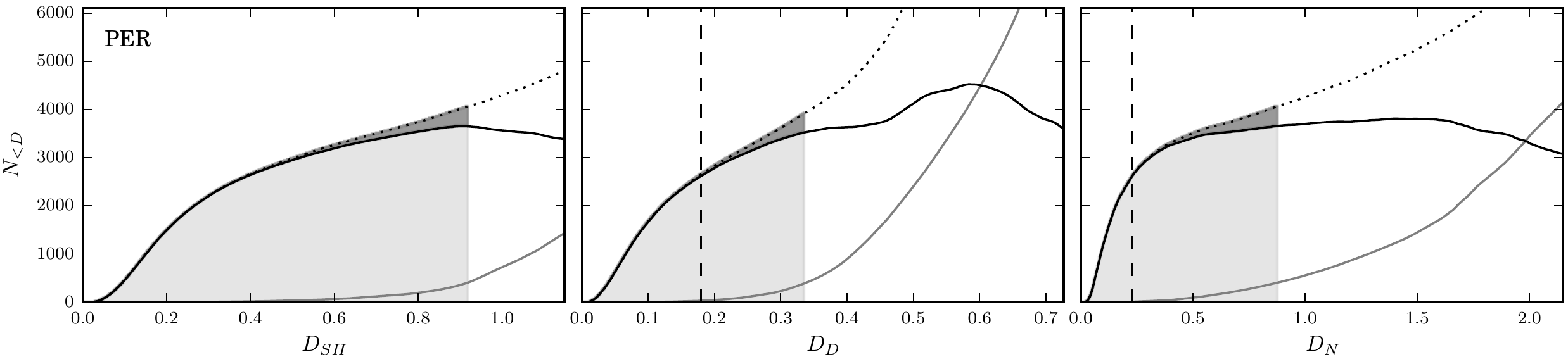} \\
\includegraphics[width=\textwidth]{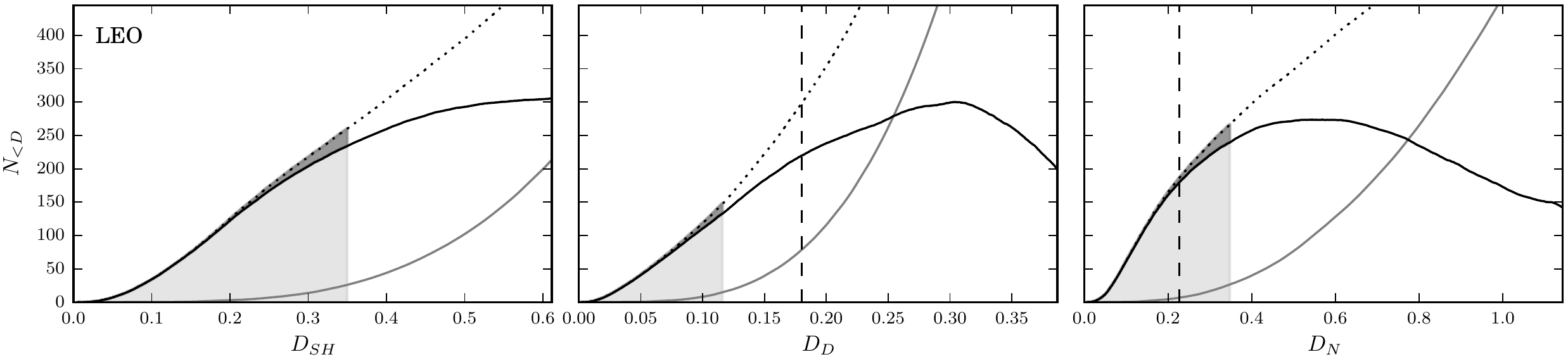} \\
\includegraphics[width=\textwidth]{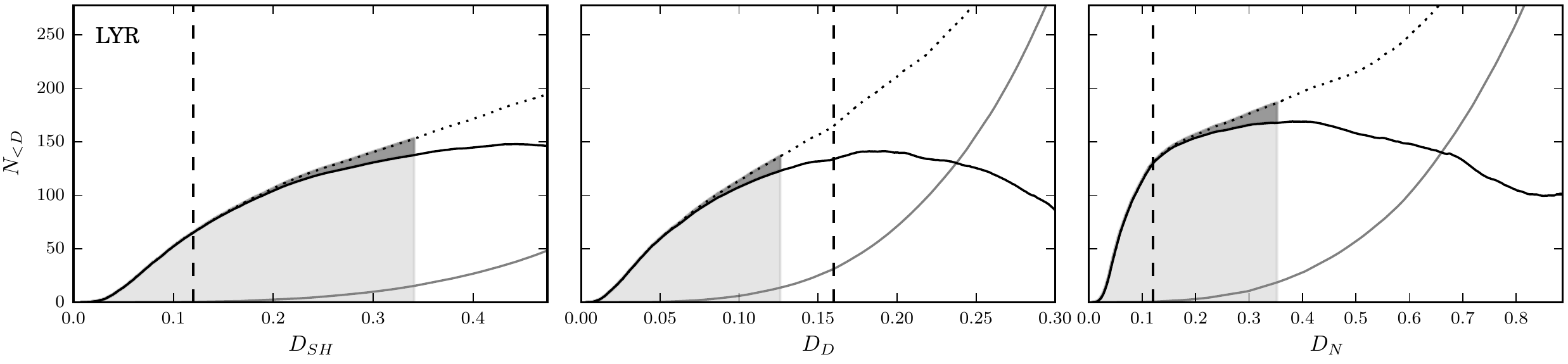} 
\caption{The cumulative distribution of orbital similarity parameters for three selected showers in our data set.  Each row corresponds to a single meteor shower, and the columns correspond to the three $D$-parameters investigated: $D_{SH}$ (left), $D_D$ (middle), and $D_N$ (right).  In each plot, the dotted black line depicts the raw CDF of each $D$-parameter computed relative to the shower orbit (Table \ref{tab:orbs}).  The solid gray line marks the average CDF for the 25 ``shower analogs."  The solid black line represents the difference between the raw $D$ cumulative distribution for the shower and the false positive $D$ cumulative distribution for the shower analogs.  The dashed blue vertical line marks Galligan's 70\% shower retrieval cutoff for each $D$-parameter, where applicable, and shaded region marks the interval in which the sporadic intrusion is less than 10\%. }
\label{fig:cdf}
\end{figure*}

Figure \ref{fig:cdf} compares $D$-parameter cumulative distributions for three selected showers.  The distribution about the nominal showers is depicted in black, while the distribution about the shower analogs, combined, appears as a solid gray line.  By comparing these distributions, we can compute the cutoff value for each $D$-parameter at which the sporadic intrusion (shaded region) reaches 10\% of the shower strength.  

Figure \ref{fig:cdf} also marks Galligan's 70\% recovery criterion for each shower and $D$-parameter, where defined \citep{2001MNRAS.327..623G}.  For many of our major showers, including the Perseids, this cutoff is significantly smaller than our recommended cutoff.  However, for the Leonids, Galligan's 70\% recovery cutoff for $D_D$ is larger than our recommended cutoff.  In this case, the blind use of Galligan's recommendation could result in including a large number of false positives.  

For smaller showers, such as the April Lyrids, Galligan's 70\% cutoff values for each $D$-parameter produce very different numbers of shower members.  These cutoffs would also produce significantly different false positive inclusion rates.  This shower is a good illustration of the value of individually characterizing the false positive rate for a given shower and $D$-parameter.

\begin{figure}
\includegraphics[width=3.3in]{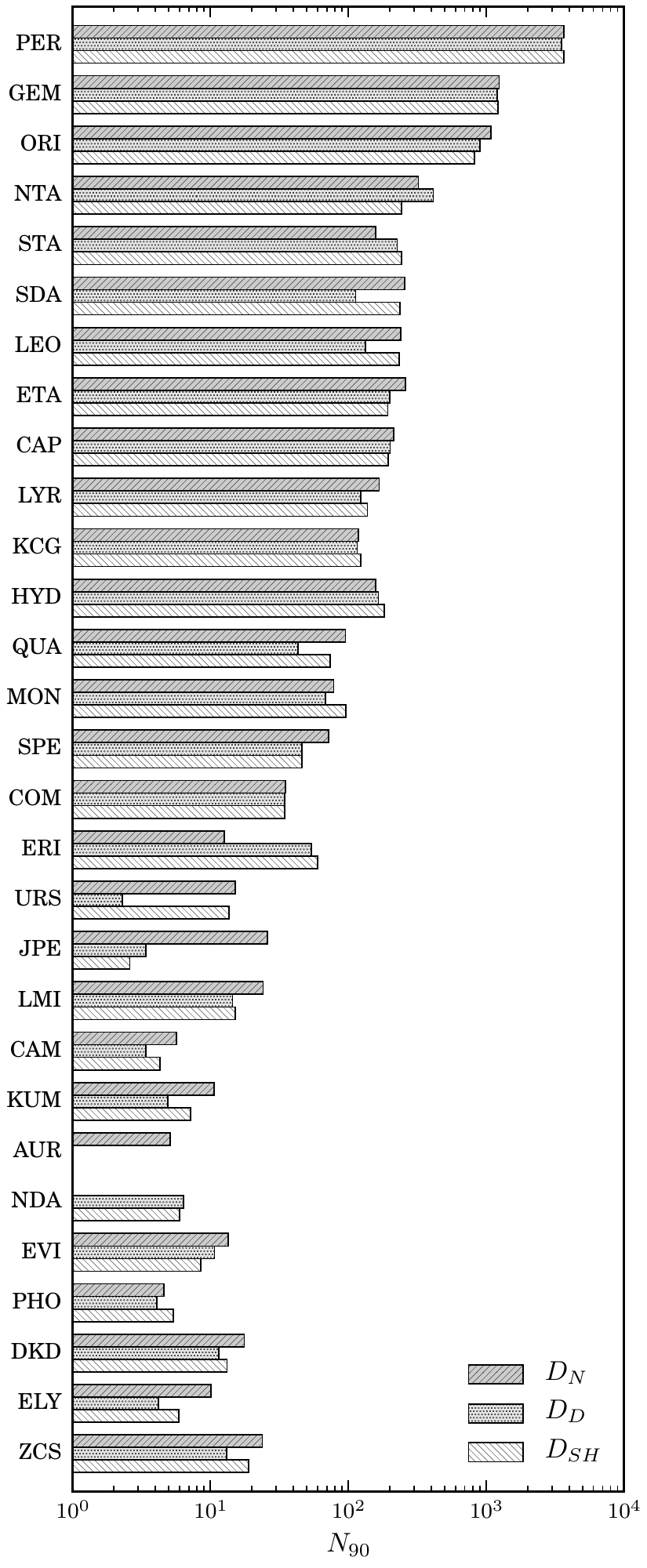}
\caption{The number of meteors that can be extracted from our data set with 90\% confidence of shower membership ($N_{90}$) for each shower using three choices of $D$-parameter.  For our largest showers, $D_N$ parameter usually retrieves the largest number of meteors.}
\label{fig:bar}
\end{figure}

In general, we agree with \cite{2001MNRAS.327..623G} that $D_N$ parameter has the best overall performance.  For most of our largest showers, $D_N$ retrieves the most shower members with 90\% confidence (see Figure \ref{fig:bar}). For smaller showers, the three parameters vary in performance.

\subsection{Shower orbits}

We have analyzed the performance of each $D$-parameter relative to either established shower parameters, or, in a few cases, relative to shower orbits computed by the authors prior to beginning this study.  By using previously measured orbits, we hope to avoid biasing our results in favor of one $D$-parameter over another.  Now that we have completed our comparison of $D$-parameter performance, we compute average shower orbits using the method described in \ref{sec:extract}.

Table \ref{tab:orbs} reports both the original reference orbits and our measured orbits.  We report our computed shower orbits only for cases where the use of our computed orbit improves the shower removal (i.e., increases $N_{90}$ for $D_N$).  In some cases, such as the Geminids, we were unable to improve on established orbits.

\begin{table*} \tiny
\centering
\begin{tabular}{ccccccccccc}
Shower & $q$ (au) & $e$ & $i$ ($^\circ$) & $\omega$ ($^\circ$) & $\Omega$ ($^\circ$) & RA ($^\circ$) & dec ($^\circ$) & $v_g$ & $\lambda_\odot$ ($^\circ$) & Source \\
\hline
Perseids & 0.95 & 0.85 & 112.47 & 150.0 & 138.74 & 45.47 & 57.55 & 57.97 & 138.74 & \cite{Brown:2010gj} \\
(PER) & 0.9519(10) & 0.835(9) & 113.0(3) & 150.04(17) & 138.33(16) & 44.8(4) & 57.0(3) & 57.971(18) & 138.33(16) & this work\vspace{0.05in} \\
Geminids & 0.1373 & 0.898 & 23.2 & 324.95 & 261.0 & 112.5 & 32.1 & 34.5 & 261 & \cite{Brown:2010gj} \vspace{0.05in} \\
Orionids & 0.5746 & 0.895 & 162.8 & 83.98 & 28.0 & 95.5 & 15.2 & 65.4 & 208 & \cite{Brown:2010gj} \\
(ORI) & 0.553(4) & 0.812(4) & 164.10(5) & 89.5(3) & 28.32(11) & 96.56(5) & 15.91(3) & 63.87(3) & 208.33(11) & this work\vspace{0.05in} \\
N.~Taurids & 0.354 & 0.8283 & 2.3 & 294.8 & 223.8 & 53.3 & 21.0 & 28.1 & 224.5 & \cite{Brown:2008iz} \vspace{0.05in} \\
S.~Taurids & 0.374 & 0.810 & 5.318 & 113.121 & 39.715 & 50.1 & 13.4 & 27.2 & 219.7 & \cite{2009JIMO...37...55S} \\
(STA) & 0.3641(12) & 0.8184(12) & 4.92(7) & 113.96(11) & 41.2(8) & 51.8(7) & 14.3(2) & 27.63(6) & 221.2(8) & this work\vspace{0.05in} \\
S.~$\delta$-Aquariids & 0.065 & 0.9726 & 30.9 & 153.9 & 306.2 & 341.0 & -16.1 & 41.1 & 126.5 &  \cite{Brown:2008iz} \\
(SDA) & 0.0901(5) & 0.96100(20) & 23.23(9) & 149.26(10) & 308.11(8) & 340.09(7) & -16.07(2) & 39.31(2) & 128.12(8) & this work\vspace{0.05in} \\
Leonids & 0.9838 & 0.61 & 162.0 & 171.11 & 237.0 & 155.1 & 21.1 & 67.3 & 237&  \cite{Brown:2010gj} \\
(LEO) & 0.98480(10) & 0.666(6) & 159.5(2) & 172.35(11) & 236.45(4) & 155.00(7) & 22.73(10) & 67.72(10) & 236.45(4) & this work\vspace{0.05in} \\
$\eta$-Lyrids & 0.537 & 0.924 & 162.9 & 91.6 & 45.1 & 338.0 & -0.7 & 64.6 & 45.5 &  \cite{Brown:2008iz}\\
(ETA) & 0.75(3) & 0.67(5) & 170.2(12) & 111(3) & 46.45(8) & 331.3(12) & -6.4(10) & 64.41(12) & 46.46(8) & this work\vspace{0.05in} \\
$\alpha$-Capricornids & 0.586 & 0.75 & 7.3 & 269.2 & 123.3 & 302.9 & -9.9 & 22.2 & 123.5 &  \cite{Brown:2008iz}\\
(CAP) & 0.586(3) & 0.7561(12) & 7.31(2) & 268.9(3) & 125.2(5) & 304.6(3) & -9.62(14) & 22.33(9) & 125.1(5) & this work\vspace{0.05in} \\
April Lyrids & 0.9149 & 0.916 & 80.0 & 215.71 & 32.0 & 272.2 & 32.6 & 46.6 & 32 & \cite{Brown:2010gj} \\
(LYR) & 0.91880(20) & 0.912(3) & 78.91(8) & 214.94(6) & 32.298(18) & 272.20(5) & 33.39(5) & 46.08(6) & 32.298(18) & this work\vspace{0.05in} \\
$\kappa$-Cygnids (KCG) & 0.975 & 0.695 & 33.399 & 204.826 & 140.702 & 285.0 & 50.10 & 21.90 & 140.70 & \cite{2009JIMO...37...55S} \vspace{0.05in}\\
$\sigma$-Hydrids (HYD) & 0.269 & 0.976 & 130.391 & 118.235 & 72.898 & 123.20 & 3.0 & 59.0 & 252.90 & \cite{2009JIMO...37...55S} \vspace{0.05in}\\
Quadrantids & 0.9746 & 0.709 & 72.4 & 168.14 & 283.0 & 231.5 & 48.5 & 41.7 & 283.0 & \cite{Brown:2010gj} \\
(QUA) & 0.98080(20) & 0.585(2) & 69.84(4) & 173.26(17) & 283.27(5) & 229.45(10) & 50.20(6) & 39.49(2) & 283.27(5) & this work\vspace{0.05in} \\
Dec.~Monocerotids & 0.1936 & 0.978 & 32.4 & 128.65 & 81.0 & 102.3 & 8.6 & 40.6 & 261.0 & \cite{Brown:2010gj} \\
(MON) & 0.174(13) & 0.960(5) & 34.0(17) & 133(2) & 80.5(2) & 103.8(10) & 8.51(13) & 40.16(8) & 260.5(2) & this work\vspace{0.05in} \\
Sep.~$\epsilon$-Perseids & 0.705 & 0.937 & 139.008 & 247.671 & 167.10 & 47.30 & 39.30 & 63.90 & 167.10 & \cite{2009JIMO...37...55S}\\
(SPE) & 0.684(4) & 0.858(5) & 137.40(15) & 252.2(6) & 167.60(7) & 47.61(16) & 39.83(4) & 62.36(12) & 167.60(7) & this work\vspace{0.05in} \\
Comae Berenicids & 0.575 & 0.993 & 135.210 & 260.461 & 268.002 & 161.50 & 30.50 & 64.0 & 268.0 & \cite{2009JIMO...37...98M}\\
(COM) & 0.5357(21) & 0.891(8) & 132.33(11) & 268.1(5) & 268.01(8) & 161.48(18) & 31.17(10) & 61.45(17) & 268.01(8) & this work\vspace{0.05in} \\
$\eta$-Eridanids (ERI) & 0.955 & 0.910 & 132.266 & 28.534 & 317.601 & 44.50 & -11.70 & 64.0 & 137.60 & \cite{2009JIMO...37...55S}\vspace{0.05in}\\
Ursids & 0.9470 & 0.961 & 55.5 & 202.53 & 270.0 & 222.1 & 74.8 & 35.6 & 270.5 & \cite{Brown:2010gj} \\
(URS) & 0.9366(6) & 0.824(3) & 51.1(2) & 206.63(19) & 269.87(17) & 220.8(3) & 77.34(19) & 32.35(12) & 269.87(17) & this work\vspace{0.05in} \\
July Pegasids & 0.649 & 1.198 & 149.230 & 250.279 & 107.997 & 347.20 & 11.10 & 68.10 & 108.0 &  \cite{2009JIMO...37...98M}\\
(JPE) & 0.573(3) & 0.941(5) & 147.83(10) & 264.2(6) & 108.28(11) & 347.43(10) & 10.93(8) & 63.60(14) & 108.28(11) & this work\vspace{0.05in} \\
Leonis Minorids & 0.638 & 0.964 & 125.916 & 105.639 & 208.901 & 158.80 & 37.10 & 61.90 & 208.90 & \cite{2009JIMO...37...55S}\\
(LMI) & 0.6181(21) & 0.893(5) & 124.0(3) & 101.4(3) & 210.10(6) & 160.41(8) & 36.97(13) & 60.30(12) & 210.10(6) & this work\vspace{0.05in} \\
May & 0.97 & 0.57 & 18.67 & 150.78 & 62.72 & 117.85 & 77.38 & 14.19 & 62.72 & this work \\
Camelopardalids & 0.9725(11) & 0.618(12) & 18.34(11) & 153.7(4) & 61.91(17) & 126.1(10) & 75.9(5) & 14.31(11) & 61.90(17) & this work\vspace{0.05in} \\
$\kappa$-Ursae Majorids & 0.99 & 0.82 & 128.65 & 188.14 & 222.80 & 144.14 & 45.78 & 63.60 & 222.80 & this work \vspace{0.05in}\\
Aurigids & 0.70 & 1.093 & 149.436 & 114.394 & 158.999 & 91.80 & 39.0 & 67.70 & 159.0 & \cite{2012JIMO...40..201M}\\
(AUR) & 0.661(3) & 0.901(4) & 147.70(13) & 105.7(4) & 158.10(10) & 90.40(10) & 39.34(6) & 64.53(7) & 158.10(10) & this work\vspace{0.05in} \\
N.~$\delta$-Aquariids & 0.0955 & 0.944 & 23.4 & 329.94 & 139.0 & 345.7 & 2.3 & 37.3 & 139 & \cite{Brown:2010gj} \\
(NDA) & 0.1046(16) & 0.9598(5) & 22.6(6) & 326.2(3) & 139.5(3) & 343.9(3) & 1.26(7) & 39.15(8) & 139.5(3) & this work\vspace{0.05in} \\
$\eta$-Virginids & 0.46 & 0.80 & 5.17 & 282.47 & 356.60 & 184.73 & 3.76 & 26.35 & 356.60 & this work\vspace{0.05in}\\
Dec.~Phoenicids & 0.98 & 0.64 & 6.84 & 8.76 & 69.10 & 6.73 & -29.20 & 9.35 & 249.08 & this work\vspace{0.05in}\\
Dec.~$\kappa$-Draconids & 0.93 & 0.85 & 73.37 & 208.42 & 250.20 & 186.0 & 70.10 & 43.40 & 250.20 & \cite{2009JIMO...37...55S}\\
(DKD) & 0.9209(7) & 0.869(5) & 72.2(3) & 210.87(14) & 250.65(3) & 184.01(19) & 71.2(2) & 43.03(11) & 250.65(3) & this work\vspace{0.05in} \\
$\eta$-Lyrids & 1.001 & 0.931 & 74.797 & 191.119 & 50.0 & 291.30 & 43.40 & 44.0 & 50.0 & \cite{2013JIMO...41..133M}\\
(ELY) & 1.0000(4) & 0.929(19) & 73.39(17) & 191.6(3) & 50.06(5) & 290.31(8) & 44.09(10) & 43.3(3) & 50.06(5) & this work\vspace{0.05in} \\
$\zeta$-Cassiopeiids & 1.0 & 0.89 & 106.21 & 164.38 & 113.11 & 5.96 & 51.01 & 56.29 & 113.11 & this work \\
(ZCS) & 0.9978(5) & 0.899(10) & 106.5(4) & 163.98(18) & 113.05(6) & 6.2(3) & 50.91(19) & 56.48(16) & 113.05(6) & this work\vspace{0.05in} \\
$\epsilon$-Geminids & 0.774 & 0.920 & 173.052 & 237.625 & 206.011 & 101.60 & 26.70 & 68.80 & 206.0 & \cite{Jenniskens:2012ur} \\
\end{tabular}  
\caption{Reference orbits used to analyze the false-positive rate for $D$-parameter-based shower association.  In a few cases, we generated our own reference orbit using meteors near peaks in an orbital element heat map (Figure \ref{fig:heatmap}) or those selected automatically by a cluster detection algorithm (Burt et al., 2014).  For each shower, we attempted to recalculate the average orbit using the method described in Section \ref{sec:extract}.  We report these results only when extraction with the average orbit produced a higher yield than extraction with the original reference orbit.}
\label{tab:orbs}
\end{table*}

\section{Conclusions}

We present a method for characterizing the expected false positive rate for shower extraction using orbital similarity criteria, or $D$-parameters.  This method involves the construction of shower ``analogs" that have the same Sun-centered ecliptic radiant and geocentric velocity as the nominal shower, but vary in solar longitude.  Shower detection occurs when the $D$-parameter distribution about the shower exceeds that about the shower analogs.

The construction of a false-positive distribution also permits the computation of the probability of shower membership as a function of $D$.  It also assists the user in selecting an orbital similarity cutoff that limits the false positive rate to within a tolerable, user-determined percentage.  We find that the traditional $D$-parameter cutoffs may fall short of or exceed what's appropriate, resulting in either needlessly throwing away shower members or including too many false positives.  We recommend instead choosing a cutoff based on the modeled sporadic intrusion, especially for small showers which may be tricky to isolate from the sporadic background.

Like \cite{2001MNRAS.327..623G}, we find that the $D_N$-criterion of \cite{1999MNRAS.304..743V} does the best overall job of isolating shower members from the sporadic background.  For most of our strongest showers, $D_N$ was able to extract the largest number of meteors while limiting the sporadic intrusion to $\le 10$\%.

\bibliographystyle{mnras}
\bibliography{fp}

\appendix

\section{Orbital similarity parameters}

This appendix presents the formulas needed to compute each of the orbital similarity parameters investigated in this paper.  These formulae are available in many papers but we repeat them here for the convenience of the reader.

\subsection{The Southworth \& Hawkins $D_{SH}$ parameter}
\label{sec:dsh}

The first-established $D$-criterion that is still in wide use is that of \cite{1963SCoA....7..261S}, which we will denote $D_{SH}$.  This parameter computes the degree of dissimilarity between two orbits as
\begin{align}
D_{SH}^2 &= (q_1 - q_2)^2  + (e_1 - e_2)^2 + \left({2 \sin{\frac{I}{2}}}\right)^2 +
\left({\frac{e_1+e_2}{2} \cdot 2 \sin{\frac{\Pi}{2}}}\right)^2 \, , \label{eq:dsh}
\end{align}
where $I$ is the angle between the two orbital planes and $\Pi$ is the angle between the two eccentricity vectors.  These angles are computed as follows:
\begin{align}
I &= \arccos{\left[{\cos{i_1} \cos{i_2} + \sin{i_1} \sin{i_2} \cos(\Omega_1 - \Omega_2)}\right]} \\
\Pi &= \omega_2 - \omega_1 \pm 2 \arcsin{\left({
\cos{\frac{i_2+i_1}{2}} \sin{\frac{\Omega_2 - \Omega_1}{2}} \sec{\frac{I}{2}}
}\right)} \, .
\end{align}
In these equations, $q_j$ is the perihelion distance of meteoroid $j$, $e_j$ is its eccentricity, $i_j$ is its inclination, $\omega_j$ is its argument of pericenter, and $\Omega_j$ is its longitude of ascending node.  The plus sign is used when $\Omega_1$ and $\Omega_2$ differ by less than $180^\circ$, and the minus sign when the difference in node is greater than $180^\circ$.  Note that for non-hyperbolic, Earth-intersecting orbits, the first two terms of Equation \ref{eq:dsh} have a maximum value of unity, while the latter two terms have a maximum value of 4.  Thus, $D_{SH} \le \sqrt{10}$.

\subsection{The Drummond $D_D$ parameter}
\label{sec:dd}

\citep{1981Icar...45..545D} revised this criterion by substituting relative for absolute differences between orbital elements and by using angles in the place of chords \citep{2001MNRAS.327..623G}.  This revised parameter is
\begin{align}
D_D^2 &=  \left({\frac{q_1-q_2}{q_1+q_2}}\right)^2 + \left({\frac{e_1-e_2}{e_1+e_2}}\right)^2 +\left({\frac{I}{180^\circ}}\right)^2 + \left({\frac{e_1+e_2}{2} \cdot \frac{\theta}{180^\circ}}\right)^2 \, .
\end{align}
Here, $\Pi$ has been replaced by $\theta$, which is defined as
\begin{align}
\theta &= \arccos [\sin \beta_1 \sin \beta_2 + \cos \beta_1 \cos \beta_2 \cos(\lambda_2 - \lambda_1)] \, .
\end{align}
Finally, $\lambda$ and $\beta$ represent the ecliptic longitude and latitude of perihelion, which can be computed as follows:
\begin{align}
\lambda &= \Omega + \arctan (\cos i \tan \omega) \\
\beta &= \arcsin (\sin i \sin \omega) \, .
\end{align}
The use of fractional differences in the first two terms ensures that each of these terms cannot exceed unity.  For non-hyperbolic orbits, $D_D \le 2$. 

\subsection{The Valsecchi et al.~$D_N$ parameter}
\label{sec:dn}

\cite{1999MNRAS.304..743V} proposed a new orbital similarity parameter, $D_N$, that is more closely linked to observable quantities: geocentric right ascension and declination ($\alpha_g$ and $\delta_g$), geocentric velocity ($v_g$), and solar longitude ($\lambda_\oplus$).  These observables are converted into variables $u$, $\theta$, and $\phi$, where $u$ is the ratio of $v_g$ to the Earth's velocity ($v_\oplus$), $\theta$ describes the angle between $\vec{v}_g$ and $\vec{v}_\oplus$, and $\phi$ describes the component of the meteoroid's geocentric velocity that is perpendicular to $\vec{v}_\oplus$.  These quantities can be calculated from observed quantities as follows.
\begin{align}
\left({
	\begin{array}{c}
	u_x \\ u_y \\ u_z
	\end{array}
}\right) &= \frac{v_g}{v_\oplus} \left({
	\begin{array}{ccc}
	\cos{\lambda_\odot} & \sin{\lambda_\odot} & 0 \\
	-\sin{\lambda_\odot} & \cos{\lambda_\odot} & 0 \\
	0 & 0 & 1
	\end{array}
}\right) \times \nonumber \\
& \left({
	\begin{array}{ccc}
	1 & 0 & 0 \\
	0 & \cos{\epsilon} & \sin{\epsilon} \\
	0 & -\sin{\epsilon} & \cos{\epsilon} 
	\end{array} 
}\right) \times \left({
	\begin{array}{c}
	-\cos{\delta_g} \cos{\alpha_g} \\ -\cos{\delta_g} \sin{\alpha_g} \\ -\sin{\delta_g}
	\end{array}
}\right)
\end{align}

\begin{align}
u &= v_g/v_\oplus \nonumber \\
\theta &= \cos^{-1}{(u_y/u)} \\
\phi &= \tan^{-1}{(u_x/u_z)} \nonumber 
\end{align}
where $\phi$ must be placed in the appropriate quadrant based on the sign of $u_x$ and $u_z$.

Finally, $D_N$ is computed from these quantities using the following set of equations:
\begin{align}
\Delta \phi_a &= 2 \sin \tfrac{1}{2}(\phi_2 - \phi_1) \nonumber \\
\Delta \phi_b &= 2 \sin \tfrac{1}{2}(\pi + \phi_2 - \phi_1) \nonumber \\
\Delta \lambda_a &= 2 \sin \tfrac{1}{2}(\lambda_{\odot, 2} - \lambda_{\odot, 1}) \nonumber \\
\Delta \lambda_b &= 2 \sin \tfrac{1}{2}(\pi + \lambda_{\odot, 2} - \lambda_{\odot, 1}) \nonumber \\
\Delta \xi^2 &= \min(w_2 \Delta \phi_a^2 + w_3 \Delta \lambda_a^2, ~ w_2 \Delta \phi_b^2 + w_3 \Delta \lambda_b^2) \nonumber \\
D_N^2 &= (u_2 - u_1)^2 + w_1 (\cos{\theta_2} - \cos{\theta_1})^2 + \Delta \xi^2 
\end{align}
Both \cite{1999MNRAS.304..743V} and \cite{2001MNRAS.327..623G} set $w_1 = w_2 = w_3 = 1$.  We experimented with setting these weights equal to that of the first term (i.e., $\sim 72/30$) but found that this tended to conflate the Orionids and the $\eta$-Aquariids with the Taurids.  Therefore, we chose to use $w_1 = w_2 = w_3 = 1$ in this work as well.

\end{document}